\documentclass[final,3p,times,twocolumn]{elsarticle}

 \biboptions{comma,sort&compress}

\usepackage{here}
 \usepackage{graphicx}
  \usepackage{epsfig}

\def\nin{\noindent}
\def\beq{\begin{equation}}
\def\eeq{\end{equation}}
\def\bea{\begin{eqnarray}}
\def\eea{\end{eqnarray}}

\def\MSbar{\relax\ifmmode\overline                        
            {\rm MS}\else{$\overline{\rm MS}${ }}\fi}     

\journal{Nuc. Phys. (Proc. Suppl.)}

\begin{document}

\begin{frontmatter}

\title{$\{\beta\}$-expansion in QCD, its conformal symmetry limit:\\
theory + applications}

 \author[label1]{A.L. Kataev  \corref{cor1}}
  \address[label1]{Institute for Nuclear Research of the Academy of Sciences of
Russia, 117312, Moscow, Russia}

\cortext[cor1]{Speaker}
\ead{kataev@ms2.inr.ac.ru}

 \author[label2]{S.V. Mikhailov}

  \address[label2]{Bogoliubov Laboratory of Theoretical
Physics, JINR, 141980 Dubna, Russia}

\ead{mikhs@theor.jinr.ru}

\begin{abstract}

\noindent

The basis of the $\{\beta\}$-expansion for the perturbative series
evaluated in the $\MSbar$ scheme for the renormalization group
invariant quantities is summarized.
Comparison with a similar representation,
used within the BLM-motivated Principle of Maximal Conformality,
is discussed.We stress that the original  $\{\beta\}$-expansion  contains a completed
list of terms rather than its PMC  analog.
The  arguments in favour of the complete $\{\beta\}$-expansion are presented.
They are  based on the relations which follow from
the power  $\beta$-function   generalization of the  Crewther relation
for the nonsinglet $\MSbar$  contributions to the Adler $D^{NS}$-function
and to the Bjorken sum rule $C^{Bjp}_{NS}$ of the polarized lepton-nucleon scattering.
The terms of the  complete $\{\beta\}$-expansion
at the $O(\alpha_s^3)$ level  for $D^{NS}$ and $C^{Bjp}_{NS}$ are presented.
These perturbative results are expressed in the PMC-type form.
The problem of applications of these expressions for phenomenological applications is summarized.

\end{abstract}

\end{frontmatter}

\section{Introduction}

\nin
The $\{\beta\}$-expansion approach, discussed here,
was originally proposed in  \cite{SM}.
The aim was  to  construct  generalizations of the
BLM approach \cite{BLM} at the levels
higher than the NNLO one
while the first method to fix the BLM-type scale for
the  RG-invariant quantities was developed in \cite{GK}.
This
$\{\beta\}$-expansion
was used  to explore  multiple power
 $\beta$-function  generalization
of the  Crewther relation in the $\MSbar$-scheme for the
nonsinglet (NS) corrections to the Adler $D$-function and
to the Bjorken sum rule of the polarized lepton-nucleon scattering
 \cite{KM1}.
Expanding this
form of the generalized Crewther relation in powers of $\alpha_s$ and keeping
 the single power of the QCD $\beta$-function only,  one can
recover the generalized Crewther relation with the
single $\beta$-function factor.
The existence of this $\MSbar$-scheme
relation  was discovered  at the $\alpha_s^3$-level
\cite{Broadhurst:1993ru} and confirmed later on in \cite{Baikov:2010je} at the
$\alpha_s^4$   order.
This  relation follows from the consideration of the AVV quark current
triangle diagram not only in the massless  quark-parton model \cite{Crewther:1972kn},
which respects conformal symmetry, but in the case, when   the insertion of
higher-order QCD  corrections to this triangle
diagram are also  taken into account \cite{Gabadadze:1995ei}.
Theoretical validity of the
generalized $\MSbar$-scheme Crewther relation,
presented as the additional term with  factored out single power of the
$\beta$-function  was studied  in \cite{Crewther:1997ux,Braun:2003rp},
where its validity in
all orders of perturbation theory was investigated.
More recently the $\{\beta\}$-expansion approach   was  explored
in \cite{KM2} in relation to  its  analog,
used in \cite{PMC1,PMC2,PMC3,PMC4,PMC5} for various applications  of the
Principle of Maximal Conformality (PMC) proposed in \cite{PMC}.
Note that the main aim of  PMC,
which is   similar to the seBLM  method in \cite{SM},
is to construct a new high-order representation of the BLM approach
by  absorbing  all terms proportional to the $\beta$-function coefficients
into the scales of each integer power of the coupling $\alpha_s$
in perturbative series for the RG-invariant quantities.
For the NS Adler
function and the  Bjorken polarized sum rule
the coefficients of these modified series should
respect the relations, which
follow from the conformal symmetry and  the original Crewther relation of
\cite{Crewther:1972kn}  (for the recent theoretical  studies of the
consequences of the conformal symmetry
in QED and QCD see \cite{Kataev:2013vua}).

\section{Comparison of the complete and incomplete $\{\beta\}$-
expansions for the $D^{NS}$-function}
\nin
Following the work \cite{KM2},  let us clarify  first  the
differences between the
complete and unique  $\{\beta\}$-expansion \cite{SM} and the incomplete one
used in the studies of \cite{PMC1,PMC2,PMC3,PMC4,PMC5}.
Within the complete $\{\beta\}$-expansion  the expression for the perturbative
coefficients of the N$^3$LO approximation of the $D^{NS}$-function
\beq
\label{Dns}
D^{NS}(a_s)=1+\sum_{n=1}^{n=4} d_{n} a_s^{n}
\eeq
is expressed  through the coefficients of the
$\beta$-function of the colour $SU(N_c)$ gauge group model
\beq
\label{bf}
\beta(a_s)=\mu^2\frac{\partial a_s}{\partial \mu^2}= -\sum_{i\geq 0}
\beta_i(N_F)a_s^{i+2}~~.
\eeq
in the following form:
\bea
\label{d1b}
d_1&=&d_1[0], \\
\label{d2b}
d_2&=&\beta_0(N_F)d_2[1]+d_2[0], \\ \nonumber
d_3&=&\beta_0^2(N_F)d_3[2]+\beta_1(N_F)d_3[0,1] \\ \label{d3b}
&&+ \underline{\beta_0(N_F)d_3[1]}+d_3[0], \\ \nonumber
d_4&=&\beta_0^3(N_F)d_4[3]+\beta_1(N_F)\beta_0(N_F)d_4[1,1 ] \\ \nonumber
&&+\beta_2(N_F)d_4[0,0,1]+
\beta_0^2d_4[2]  \\ \label{d4b}
&&+ \underline{\beta_1(N_F)d_4[0,1]}+\underline{\beta_0(N_F)d_4[1]}+d_4[0],
\eea
where $N_F$ is the number
of fermion  flavours and the {\it underlined} terms were neglected
in similar expansions used in \cite{PMC1,PMC2,PMC3,PMC4,PMC5}.
The reason of neglecting them  is related to the fact  that the authors of these works
define their  $\{\beta\}$-expansions from  the traditional  expressions
f0r $d_i$ coefficients  expanded  in powers of $N_F$, namely
\bea
\label{1}
d_1&=&N_F^0d_1 \\
\label{2}
d_2&=&N_Fd_{21}+d_{20} \\
\label{3}
d_3&=&N_F^2d_{32}+N_Fd_{31}+d_{30}      \\
\label{4}
d_4&=&N_F^3d_{43}+N_F^2d_{42}+N_Fd_{41}+d_{40}~~~.
\eea
However, it is already known that to formulate the  generalized BLM
approach at the NNLO using  Eqs.(7-9),
it is necessary to take into account some  extra
information \cite{GK,SM}.
Within the   approach of   \cite{SM} extra terms,
which allow one to obtain the  complete $\{\beta\}$ expansion of $d_3$ in
Eq.(5), are  the analytical contributions of the multiplet of
light gluinos $n_{\tilde{g}}$ to the $O(\alpha_s^3)$ approximations
of the $D^{NS}$-function
\cite{Chetyrkin:1996ez} and to the $\beta$-function of the
$SU(N_c)$ group, which was  evaluated in the $\MSbar$-scheme
at the three-loop level in \cite{Clavelli:1996pz}.
The application of  new  degrees of freedom $n_{\tilde{g}}$
in both $D^{NS}(a_s)$ and $\beta(a_s)$-functions
at the $O(\alpha_s^3)$ level allowed
splitting in the   $\{\beta\}$-expansion of  the
 $\beta_1(N_F)$ and $\beta_0(N_F)$-dependent  terms, both
 contributing to the  $N_F$-term of $d_3$ in  Eq.(\ref{3}).
We have thus  obtained \cite{SM} the elements in the RHS of Eqs.(\ref{d1b}-\ref{d3b}),
which define the following
new matrix representation for $D^{NS}$:
\beq
\label{matrix}
D^{NS}(Q^2)=1+\sum_{n\geq 1}\sum_{l}a_s^{n}(Q^2)d_{n}[l]B_l(N_F)
\eeq
In Eq.(\ref{matrix}),
the  $B_l(N_F)$ factors  are the products of the $\beta$-function coefficients
of  Eqs.(\ref{d1b}-\ref{d4b}),
$d_n(N_F)=d_{n}[l]B_l(N_F)$ are the $N_F$-dependent coefficients  in
Eqs.(\ref{1}-\ref{4})  while the elements $d_{n}[l]$ do not depend on the
numbers of flavours $N_F$.
Note that  in view of the absence of an analytical
result for the gluino contributions to $D^{NS}$ at  the $O(\alpha_s^4)$ level,
we are unable to get most part of the terms in the
$\{\beta\}$-expansion of $d_4$ in Eq.(\ref{d4b}).
Indeed, only the leading
$\beta_0^{3}(N_F)d_4[3]$-contribution  is known from analytical calculations of
\cite{Broadhurst:1993ru}. This information was already used in the all-order
generalization of the BLM approach of \cite{Beneke:1994qe},
based on absorbing into the BLM scale these renormalon-type terms only.
Since we are interested in the resummation of all $\{\beta\}$-dependent terms,
we will consider  only certain expressions at the $O(\alpha_s^3)$-level.

In the  $\MSbar$-scheme, at this order of perturbation theory
the elements of the $\{\beta\}$-expansion
for $D^{NS}$
have the following analytic
form \cite{SM}:
\bea
\label{d10a}
d_1[0]&=&\frac{3}{4}~C_F \\ \label{d21a}
d_2[1]&=&\bigg(\frac{33}{8}-3\zeta_3\bigg)~C_F \\
\label{d20a}
d_2[0]&=&-\frac{3}{32}~C_F^2+\frac{1}{16}~C_FC_A \\
\label{d32a}
d_3[2]&=&\bigg(\frac{151}{6}-19\zeta_3\bigg)~C_F \\
\label{d301a}
d_3[0,1]&=&\bigg(\frac{101}{16}-6\zeta_3\bigg)~C_F \\
\label{d31a}
d_3[1]&=&\bigg(-\frac{27}{8}-\frac{39}{4}\zeta_3+15\zeta_5\bigg)~
C_F^2 \\ \nonumber
&&-\bigg(\frac{9}{64}-5\zeta_3+\frac{5}{2}\zeta_5\bigg)~C_FC_A \\
\label{d30a}
d_3[0]&=&-\frac{69}{128}~C_F^3+\frac{71}{64}~C_F^2C_A \\ \nonumber
&&+
\bigg(\frac{523}{768}-\frac{27}{8}\zeta_3\bigg)~C_FC_A^2
\eea
Note once more that in the PMC  studies of
\cite{PMC1,PMC2,PMC3,PMC4,PMC5} an analog of the $d_3[1]$-term was
absent ( or nullified). Therefore,  the remaining
$\MSbar$-scheme
contributions to Eq.(\ref{d3b})
will differ from the ones presented in
Eqs.(\ref{d301a}) and (\ref{d30a}).

\section{
The $\bar{MS}$ -scheme generalized
Crewther relation and the  $\{ \beta \}$-expansion for
the  Bjorken polarized sum rule }
\nin
The Bjorken polarized sum rule, which is still
interesting for phenomenological studies \cite{Khandramai:2011zd,Deur:2014vea},
is  defined as
\beq
S_{Bjp}=\int_0^1 g_1^{lp-ln}(x,Q^2)dx= \frac{g_A}{6}C^{Bjp}(a_s)   .
\eeq
Its coefficient function $C^{Bjp}$ contains the NS and singlet (SI) contributions
\beq
C^{Bjp}(a_s)=C^{Bjp}_{NS}(a_s)+C^{Bjp}_{SI}(a_s) .
\eeq
The existence  of the SI  term at the $O(\alpha_s^4)$ level was
demonstrated in \cite{Larin:2013yba}, though its analytical
expression is not yet fixed  by  direct diagram-by-diagram
calculations.

The $\{\beta\}$-expansion pattern is now applied to
the coefficients $c_{n}$ of the
perturbative approximation  for
$C^{Bjp}_{NS}(a_s)$ \cite{Baikov:2010je}:
\beq
C^{Bjp}_{NS}(a_s)=1+\sum_{n=1}^{n=4} c_n a_s^{n}\,,
\eeq

\bea
\label{c1b}
c_1&=&c_1[0] \\
\label{c2b}
c_2&=&\beta_0(N_F)c_2[1]+c_2[0] \\ \nonumber
c_3&=&\beta_0^2(N_F)c_3[2]+\beta_1(N_F)c_3[0,1] \\ \label{c3b}
&&+ \underline{\beta_0(N_F)c_3[1]}+c_3[0] \\ \nonumber
c_4&=&\beta_0^3(N_F)c_4[3]+\beta_1(N_F)\beta_0(N_F)c_4[1,1 ] \\ \nonumber
&&\!\!\!\!+\beta_2(N_F)c_4[0,0,1]+
\beta_0^2c_4[2]  \\ \label{c4b}
&&\!\!\!\!+ \underline{\beta_1(N_F)c_4[0,1]}+\underline{\beta_0(N_F)c_4[1]}+c_4[0].
\eea
These   coefficients are
related to similar ones,  which enter into the  $\{\beta\}$-expansion
of the perturbative series  for   $D^{NS}$ through the
multiple power  $\beta$-function form of the    generalization
of the  Crewther relation \cite{KM1}.
Note that the application of
the $\MSbar$-scheme generalization of the Crewther relation,
considered in  \cite{PMC4}, gives the $\{\beta\}$- expanded expressions for the
coefficients of the $C^{Bjp}_{NS}(a_s)$-series
without the terms underlined in Eqs.(\ref{c3b},\ref{c4b}).

We will show  that the absence of
these    terms in the studies of  \cite{PMC1,PMC2,PMC3,PMC4,PMC5}
{\it contradicts} the existing analytical $\MSbar$-scheme
$O(\alpha_s^3)$ results  for the $D^{NS}(Q^2)$ function
\cite{Gorishnii:1990vf,Surguladze:1990tg,Chetyrkin:1996ez}
and the $\MSbar$-scheme generalization of the Crewther relation
\cite{Broadhurst:1993ru,Baikov:2010je} written down in the
 multiple power  $\beta$-function representation of  \cite{KM1}.
 This part of the talk follows from the studies of \cite{KM2}.

The
$\MSbar$-scheme single $\beta$-function expression
for the generalized   Crewther relation has the following
form:
\beq
\label{Crew1}
D^{NS}(a_s)C^{Bjp}_{NS}(a_s)=1+\frac{\beta(a_s)}{a_s}K(a_s)~~.
\eeq
Here  $K(a_s)=K_1a_s+K_2a_s^2+K_3a_s^3 +O(a_s^4)$ is the polynomial, where
the known coefficient $K_1$
depends on the $SU(N_c)$  Casimir operator $C_F$ while the coefficients  $K_2$ and $K_3$
 also known analytically depend on $C_F$, $C_A$,
$T_F$ and $N_F$.
This form, originally discovered  in \cite{Broadhurst:1993ru}
at the $O(a_s^3)$ level, was recently confirmed by direct $O(a_s^4)$
calculations of $D^{NS}$ and $C^{Bjp}_{NS}$ performed in the colour
$SU(N_c)$ gauge  group theory \cite{Baikov:2010je}.
In \cite{KM1}, it was demonstrated that
Eq.(\ref{Crew1}) can be rewritten as
\bea
\label{CrwPol}
\!\!\!\!D^{NS}(a_s)C^{Bjp}_{NS}(a_s)\!\!=\!\!
1+\frac{\beta(a_s)}{a_s}\sum_{n=1}^{3} \bigg(\frac{\beta(a_s)}{a_s}\bigg)^{n-1}\!\!\!\!\!\! P_n(a_s)&& \\
=1+\sum_{n\geq 1}\sum_{r\geq 1}P_n^{r}[k,m]C_F^{k}C_A^{m}a_s^{r}~~~&&\nonumber
\eea
where $k+m=r$ and the coefficients $P_n^{r}[r,m]$ contain rational fractions and
Riemann $\zeta$-functions of odd arguments.
In Eq.(\ref{CrwPol}), the known  coefficients of the polynomials
$P_n(a_s)$ \textit{ do not depend} on $T_F N_F$ (for a more obvious
clarification of this property see the second expression in   Eq.(\ref{CrwPol}))
and are expressed by means of the coefficients  of the  $\{\beta\}$-expansion as
\bea
P_1(a_s)&=&-a_s\bigg[c_2[1]+d_2[1]\nonumber \\
&&+ a_s\bigg( c_3[1]+d_3[1]+d_1(c_2[1]-d_2[1])\bigg) \nonumber \\
&&+a_s^2\bigg(c_4[1]+d_4[1]+d_1(c_3[1]+d_3[1])\bigg)\nonumber \\
&&+d_2[0]c_2[1]+d_2[1]c_2[0]\bigg] \label{P1}\\
\nonumber
P_2(a_s)&=&a_s\bigg[c_3[2]+d_3[2]+a_s\Big(
c_4[2]+d_4[2] \\
&&-d_1(c_3[2]-d_3[2])\Big)\bigg] \label{P2} \\
P_3(a_s)&=&-a_s\bigg[c_4[3]+d_4[3]\bigg]\label{P3}
\eea
Using
Eq.(\ref{CrwPol}) the  following  relations between the
elements of the  $\{\beta\}$-expansions of Eqs.(\ref{d1b}-\ref{d4b}) and
Eqs.(\ref{c1b}-\ref{c4b}) were  obtained \cite{KM1}:
\bea
&&0=c_n[0]+d_n[0]+\sum_{l=1}^{n-1}d_l[0]c_{n-l}[0] \label{Crew} \\
&&C_F\bigg(-\frac{21}{8}+3\zeta_3\bigg) =
-c_2[1]-d_2[1]= \nonumber \\
&&-c_3[0,1]-d_3[0,1]=\ldots\label{P11}  \\ \label{P21}
&&\!\!\!\!\!\!-c_3[1]-d_3[1]-d_1(c_2[1]-d_2[1])= \\ \nonumber
&&\!\!\!\!\!\! -c_4[0,1]-d_4[0,1]-d_1(c_3[0,1]-d_3[0,1])=\ldots
\eea
Using Eqs.(\ref{d10a}-\ref{d30a})
for $d_n[l]$
and solving then either Eq.(\ref{Crew}) (initial Crewther relation \cite{Crewther:1972kn})
or Eq.(\ref{P11}) (from \cite{KM1}) we got
the elements of the  $\{\beta\}$-expansion
for $C^{Bjp}_{NS}$ at the $O(a_s^3)$ level:
\bea
\label{c10a}
c_1[0]&=&-\frac{3}{4}~C_F \\ \label{c21a}
c_2[1]&=&-\frac{3}{2}~C_F \\
\label{c20a}
c_2[0]&=&\frac{21}{32}~C_F^2-\frac{1}{16}~C_FC_A \\
\label{c32a}
c_3[2]&=&-\frac{151}{24}~C_F \\
\label{c301a}
c_3[0,1]&=&\bigg(-\frac{59}{16}+3\zeta_3\bigg)~C_F \\
\label{c31a}
c_3[1]&=&\bigg(\frac{83}{24}-\zeta_3\bigg)~
C_F^2 \\ \nonumber
&&+\bigg(\frac{215}{64}-6\zeta_3+\frac{5}{2}\zeta_5\bigg)~C_FC_A \\
\label{c30a}
c_3[0]&=&-\frac{3}{128}~C_F^3-\frac{65}{64}~C_F^2C_A \\ \nonumber
&&-
\bigg(\frac{523}{768}-\frac{27}{8}\zeta_3\bigg)~C_FC_A^2~~.
\eea
Apart from the presented above analytical expressions,
we come to the definite  theoretical conclusions.
First, we note
 that  Eq.(\ref{Crew})  is the consequence of the
Crewther relation \cite{Crewther:1972kn} and of  the  conformal symmetry.
However, it does not give us a possibility to say anything about  the {\it scheme-independence}
of the coefficients $d_i[0]$ and $c_j[0]$ even within the  MS-like schemes.
We can only conclude that the coefficients of the PMC series are
{\it scheme-dependent} but obey {\it the scheme-independent}
relation, which {\it follows from the conformal symmetry}.

Second, the chain  of Eqs.(\ref{P21}) clearly demonstrates,
that the   $\MSbar$ analytical calculations of $D^{NS}$, $C_{Bjp}$ and
the $\MSbar$-scheme generalizations of the  Crewther relations
of Eq.(\ref{Crew1}), (\ref{CrwPol}) do not allow one to {\it neglect
(or nullify)} the  terms $d_3[1]$, $d_4[0,1]$ in the $\{\beta\}$-expansion
of the  coefficients
Eq.(\ref{d3b}) and Eq.(\ref{d4b}) of   the  $D^{NS}$
RG-invariant function  and of their analogs $c_3[1]$, $c_4[0,1]$
in the perturbative expansion of the Bjorken polarized  sum rule.
Indeed, their absence  contradicts the analytical  result on the LHS
of Eq.(\ref{P21}), obtained in \cite{KM1} from
the $\MSbar$-scheme generalization of the Crewther
relation.
In view of this,
the theoretical and phenomenological studies of the works
\cite{PMC1,PMC2,PMC3,PMC4,PMC5}, where the discussed above nonzero terms
were neglected, should be reconsidered. This was done in part in \cite{KM2}
and we  will summarize   below the concrete foundations of this work.

\section{The definition of the scale-fixing prescription }
\nin
To define the generalized BLM approach within the {\it  complete} and
{\it  unique} $\{\beta\}$-expansion approach of \cite{SM},
one should absorb all $\beta$-dependent terms  of  the
$\{\beta\}$-expanded coefficients into the scales of the coupling constants.
Following the study of \cite{KM2}, let us absorb
all    $\beta$-dependent terms  of  the  coefficients
in Eqs.(\ref{d2b},\ref{d3b})
into the new scales  of the related perturbative expansions of the
$D^{NS}$-function and $C^{Bjp}_{NS}$ RG-invariant quantities.
Using   the solution of the RG-equation in Eq.(\ref{bf}) we  reexpress
the QCD running coupling constant  $a_s(\mu^2)$
in terms of the new one   $a_s^{'}=a_s(\mu^{'2})$ in the following form
considered in \cite{KM2}, namely:
\bea
\label{redefin}
a_s(t)&=&a^{'}-\beta(a_s^{'})\frac{\Delta}{1!}
 +\beta(a_s^{'})\partial_{a_s^{'}}\beta(a_s^{'})\frac{\Delta^2}{2!} \\ \nonumber
&+&\beta(a_s^{'})\partial_{a_s^{'}}(\beta(a_s^{'})\partial_{a_s^{'}}\beta(a_s^{'}))
\frac{\Delta^3}{3!}+\dots
\eea
The term $\Delta$ defines the  shift of the scales as
\beq
\Delta=t-t^{'}=ln(\mu^2/\mu^{'2})
\eeq
where $t=ln(Q^2/\mu^2)$ and $t^{'}=ln(Q^2/\mu^{'2})$.
To define  all-order  generalization of the BLM approach proposed in \cite{SM},
it is necessary to introduce the coupling constant dependent shift
\beq
\label{shift}
\Delta=\Delta(a_s^{'})=\Delta_0+a_s^{'}\beta_0\Delta_1+(a_s^{'}\beta_0)^2\Delta_2+
\dots,
\eeq
where the coupling constant dependence of this shift was first introduced
in \cite{GK} in the process of the first
formulation of the  NNLO generalization of the BLM approach.
Fixing now $Q^2=\mu^{'2}$
we obtain  the $\{\beta\}$-expansions of the  transformed to the new scale
coefficients
$d_n^{'}$ of the perturbative expressions for the $D^{NS}$-function.
They have the following form \cite{KM2}:
\bea
\label{d1bp}
d_1^{'}&=&d_1[0] \\
d_2^{'}&=&\beta_0\,d_2[1]+ d_2[0] -\beta_0\Delta_{0} \label{d2bp}\\
d_3^{'}&=&\beta_0^2 (d_3[2]- 2d_2[1]\Delta_{0} +
\Delta_{0}^2)  \nonumber\\
&&\!\!\!\!+
\beta_1(d_3[0,1]-\Delta_{0}) +\beta_0(d_3[1] -2d_2[0]\Delta_{0}) \nonumber \\
&&\!\!\!\!+d_3[0]+\beta_0^2\Delta_{1} \label{d3bp}
\eea
For the sake of  generality, we  also present the  expression for
the fourth term, which due to still incomplete analytical information
on its $\beta$-expansion can not be involved in the concrete
numerical studies
\bea
d_4^{'}&=&\beta_0^3(d_4[3] -3d_3[2]\Delta_{0}+ 3d_2[1]\Delta_{0}^2
-\Delta_{0}^3 \nonumber\\ \nonumber
&&- 2(\Delta_{0}-d_2[1])\Delta_{1}) \\ \nonumber
&&+\beta_1\beta_0(d_4[1,1]-(3d_3[0,1]+2d_2[1])\Delta_{0} \\ \nonumber
&&+\frac{5}2\Delta_{0}^2-\Delta_{1})
+\beta_2(d_4[0,0,1]- \Delta_{0}) \\ \nonumber
&&+\beta_0^2(d_4[2] -3d_3[1]\Delta_{0} \\ \nonumber
&&+3d_2[0]\Delta_{0}^2-2d_2[0]\Delta_{1}) \\ \nonumber
&&+\beta_1(d_4[0,1]-2d_2[0]\Delta_{0}) \\ \nonumber
&&+\beta_0(d_4[1]-3d_3[0]\Delta_{0}) \\
&&+ d_4[0]-\beta_0^3\Delta_{2}\label{d4bp}
\eea
The general idea of \cite{KM2}
is to absorb all $\{\beta\}$-dependent terms in
Eqs.(\ref{d2bp}-\ref{d4bp}), including the ones omitted in
\cite{PMC1,PMC2,PMC3,PMC4,PMC5}, namely the terms
proportional to  $d_3[1]$, $d_4[0,1]$ and $d_4[1]$.
Then, we accumulate these terms in ``shift'' coefficients
 $\Delta_0$, $\Delta_1$ and $\Delta_2$, which defines the
new BLM (PMC-type) scales of $D^{NS}(a'_s)$.
We will present here the results of  application of this procedure
at the $O(a_s^3)$ level only,
where all coefficients of the $\beta$-dependent terms are already determined,
see Eqs.(\ref{d21a}-\ref{d31a}) and Eqs.(\ref{c21a}-\ref{c31a}).
\section{The concrete analytical and numerical $O(\alpha_s^3)$ studies}
\nin
Following the studies \cite{KM2}, and
 solving Eqs.(\ref{d2bp},\ref{d3bp}) with respect
 to $\Delta_{0},\Delta_{1}$ and similar
expressions for $C^{Bjp}_{NS}$
in the case of ordinary QCD, we arrive at the concrete  analytical and
numerical results for the parameters in the defined in Eq.(\ref{shift})
scale $\Delta$ of
the  PMC-type BLM generalization
of $O(a_s^3)$ approximations for   $D^{NS}$ and  $C^{Bjp}_{NS}$:
\bea
\label{DeltaD0}
\Delta_0&=&d_2[1]= \frac{11}{2}-4\zeta_3=0.69177\\  \label{DeltaC0}
\bar{\Delta}_0&=&c_2[1]=-2 \\ \label{DD1n}
\Delta_1&=&\frac{1}{\beta_0^2}\big[\beta_0^2(d_3[2]-d_2[1]^2) \nonumber \\
&&+\beta_1(d_3[0,1]-d_2[1]) \nonumber\\
&&+\beta_0(d_3[1]-2d_2[0]d_2[1])\big] \label{CD1n}\\
\bar{\Delta}_1&=&\frac{1}{\beta_0^2}\big[\beta_0^2(c_3[2]-c_2[1]^2) \nonumber \\
&&+\beta_1(c_3[0,1]-c_2[1]) \nonumber\\
&&+\beta_0(c_3[1]-2c_2[0]c_2[1])\big] \,.
\eea
Eqs.(\ref{DD1n},\ref{CD1n}) contain noticeable contributions
of the  terms omitted  in \cite{PMC1,PMC2,PMC3,PMC4,PMC5},
that are proportional to $d_3[1]$ and $c_3[1]$.
Note  that for  normalization, used here, we have
$a_s=\alpha/\pi$,
$\beta_0=11/4-N_F/6$ and $\beta_1= 51/8-19N_F/24$.
The approximate  numerical  expressions for
the coefficients of the $\{\beta\}$-expansion
for the $D^{NS}$ and $C^{Bjp}_{NS}$ RG-invariant functions  read
\bea
\label{d10n}
d_1[0]&=&1~~~~~~~~~~~~~~~~~~~c_1[0]=-1 \\
d_2[1]& \approx &0.69~~~~~~~~~~c_2[1]=-2 \\
d_2[0]&\approx &0.083~~~~~~~~~~c_2[0]\approx0.917 \\
d_3[2]&\approx &3.105~~~~~~~~~~c_3[2] \approx -6.39\\
d_3[0,1]&\approx&-1.2~~~~~~~c_3[0,1]\approx -0.108 \\
d_3[1]&\approx&13.926~~~~~~~~~c_3[1]\approx -10 \\
d_3[0]&\approx&-35.87~~~~~~~c_3[0] \approx 35.03
\eea
We note poor convergence of the $O(\alpha_s^3)$ approximations of the
perturbative series constructed from  the respective conformal symmetry
coefficients $d_n[0]$ and $c_n[0]$.
Indeed, the concrete result for  the normalized $NS$ contribution to the $e^+e^-$ R-ratio,
which is related to the $D^{NS}(Q^2)$-function, has  the following form \cite{KM2}:
\bea
\label{DNSI}
R^{NS}(s)&=&1+a_s(s_{\rm PMC})+0.0833~a_s^2(s_{\rm PMC}) \\ \nonumber
&-& 35.872~a_s^3(s_{\rm PMC})+O(a_s^4)
\eea
where the  scale is defined through the solution of Eq.(\ref{DeltaD0})
and Eq.(\ref{DD1n}) . We will present it for $N_F=3$ numbers of active
flavours with $\beta_0=2.25$ and $\beta_1=4$.
It has the following expression:
\beq
\label{QD}
s_{\rm PMC}=s \cdot \exp[-0.69-3.98\beta_0a_s^{'}(s)]
\eeq
 Note that at the NLO we reproduce the standard  BLM coefficient,
which is rather small.
However, the value of the NNLO coefficient is negative and huge.
A similar feature was already observed in the case
of applications of the first generalization of the BLM approach
based on resummation of the  $N_F$-dependent corrections \cite{GK}.
This result of \cite{GK} was confirmed in \cite{PMC1}.
Applying the same procedure to
$C^{Bjp}_{NS}$ in \cite{KM2}  we  got
\bea
C^{Bjp}_{NS}(Q^2)\!\!\!\!\!&=&\!\!\!\!\!1-a_s(Q^2_{\rm PMC})+0.917 a_s(Q^2_{\rm PMC}) \\ \nonumber
&&+ 35.03 a_s^2(Q^2_{\rm PMC})+O(a_s^3)
\eea
where
\beq
Q^2_{\rm PMC}=Q^2\cdot \exp[-2-7.32\beta_0a_s^{'}(Q^2)]
\eeq
Similar results were previously obtained at the NNLO for the Bjorken
polarized sum rule within the procedure of \cite{GK} in \cite{Kataev:1992jm}.
\section{Conclusion}
\nin
We would like to emphasize that the proposed
in \cite{SM} and used later on in \cite{KM1,KM2}
$\{\beta\}$-expansion approach allows one to fix the special  terms
 $d_3[0]$ and $c_3[0]$ of  the  $\MSbar$-scheme series
for the   $e^+e^{-}$ characteristic  $R^{NS}(s)$ and for
the Bjorken polarized sum rule.
They  satisfy the  relations, which  follow
from the {\it conformal symmetry}.
However, leaving only these terms in the $O(a_s^3)$ approximations
for the special generalizations of the BLM procedure one gets
huge coefficients  related with $O(a_s^3)$ level.
In view of this, the direct  applications of theoretically interesting PMC-type
(or seBLM-type)   approximations
in the phenomenological studies should be treated with care.
\section*{Acknowledgements}
\nin
The work was supported in part  by the
Russian Foundation for Basic Research, Grant No.
14-01-00647.  The
work of MS was also supported by the BelRFFI--JINR, grant F14D-007.
One of us (ALK) wishes to thank S.Narison for invitation to the QCD-2014
and hospitality in Montpelier.
At the final stage of this work, the studies of the consequences
of the multiple $\beta$-function representation
of the generalized Crewther relation by ALK were  supoported in part
by RSCF,  Grant N 14-22-00161.

\end{document}